\documentclass[aps,prb,showpacs,twocolumn,superscriptaddress]{revtex4}
\usepackage{amsmath,graphics,amssymb}
\usepackage[next]{inputenc}
\usepackage[dvips]{epsfig}
\def\be{\begin{equation}}
\def\ee{\end{equation}}

\def\bi{\begin{itemize}}
\def\ei{\end{itemize}}
\def\bn{\begin{enumerate}}
\def\en{\end{enumerate}}
\def\bea{\begin{eqnarray}}
\def\eea{\end{eqnarray}}
\def\no{\nonumber}
\def\ba{\begin{array}}
\def\ea{\end{array}}
\def\bd{\begin{displaymath}}
\def\ed{\end{displaymath}}

\begin{document}
\title{Phase Diagram and Entanglement of Ising Model With Dzyaloshinskii-Moriya Interaction (IDM)}

\author{R. Jafari}
\affiliation{Institute for Advanced Studies in Basic Sciences,
Zanjan 45195-1159, Iran} \affiliation{School of physics, IPM (Institute for Studies in
Theoretical Physics and Mathematics), \\P. O. Box: 19395-5531, Tehran, Iran}
\author{M. Kargarian}
\affiliation{Physics Department, Sharif University of Technology,
Tehran 11155-9161, Iran}
\author{A. Langari}
\affiliation{Physics Department, Sharif University of Technology,
Tehran 11155-9161, Iran} \email[]{langari@sharif.edu}
\author{M. Siahatgar}
\affiliation{Physics Department, Sharif University of Technology,
Tehran 11155-9161, Iran} \email[]{langari@sharif.edu}
\homepage[]{http://spin.cscm.ir}

\begin{abstract}
We have studied the phase diagram and entanglement of the one
dimensional Ising model with Dzyaloshinskii-Moriya (DM) interaction.
We have applied the quantum renormalization group (QRG) approach to
get the stable fixed points, critical point and the scaling of
coupling constants. This model has two phases, antiferromagnetic and
saturated chiral ones. We have shown that the staggered
magnetization is the order parameter of the system and DM
interaction produces the chiral order in both phases. We have also
implemented the exact diagonalization (Lanczos) method to calculate
the static structure factors. The divergence of structure factor at
the ordering momentum as the size of systems goes to infinity
defines the critical point of the model. Moreover, we have analyzed
the relevance of the entanglement in the model which allows us to
shed insight on how the critical point is touched as the size of the
system becomes large. Nonanalytic behavior of entanglement and
finite size scaling have been analyzed which is tightly connected to
the critical properties of the model. It is also suggested that a
spin-fluid phase has a chiral order in terms of new spin operators
which are defined by a nonlocal transformation.

\end{abstract}
\date{\today}


\maketitle
%%%%%%%%%%%%%%%%%%%%%%%%%%%%%%%%%%%%%%%%%%%%%%%%%%%%%%%%%%%%%%%%%%%%%
\section{Introduction \label{introduction}}
At zero temperature, the properties of quantum many-body system is
dictated by the structure of its ground state. The degree of
complexity of this structure is different for various systems. It
ranges from exceptionally simple case (when a strong magnetic
field aligns all the spins of a ferromagnet along the field
direction, producing a product or unentangled state) to more
intricate situation where entanglement pervades the ground state
of system. Thus, entanglement appears naturally in low temperature
quantum many body systems, and it is at the core of relevant
quantum phenomena, such as superconductivity\cite{Bardeen},
quantum Hall effect\cite{Laughlin}, and other quantum phase
transitions\cite{Sachdev}. Quantum phase transitions have been one
of the most interesting topics of strongly correlated systems
during the last decade. It is basically a phase transition at zero
temperature where the quantum fluctuations play the dominant role
\cite{vojta}. Suppression of the thermal fluctuations at zero
temperature introduces the ground state as the representative of
the system. The properties of the ground state may be changed
drastically shown as a non-analytic behavior of a physical
quantity by reaching the quantum critical point (QCP). This can be
obtained by tunning a parameter in the Hamiltonian, such as the
magnetic field or the amount of disorder. The ground state of a
typical quantum many body system consists of a superposition of a
huge number of product states. Understanding this structure is
equivalent to establishing how subsystems are interrelated, which
in turn is what determines many of the relevant properties of the
system. In this sense, the study of entanglement offers an
attractive theoretical framework from which one may be able to go
beyond customary approaches to the physics of quantum collective
phenomena\cite{Preskill}.

Recently some novel magnetic properties
%antiferromagnetic (AF) systems
were discovered in a variety of quasi-one dimensional materials
that are known to belong to the class of Dzyaloshinskii-Moriya
(DM)
$\big(\overrightarrow{D}.(\overrightarrow{S_{i}}\times\overrightarrow{S_{j}})\big)$
magnet to explain helical magnetic structures. The relevance of
antisymmetric superexchange interactions in spin Hamiltonians
which describe quantum antiferromagnetic (AF) systems was
introduced phenomenologically by
Dzyaloshinskii\cite{Dzyaloshinskii}. Moriya showed later, that
such interactions arises naturally in the perturbation theory due
to the spin-orbit coupling in magnetic systems with low
symmetry\cite{Moriya}. Some AF systems are expected to be
described by DM interaction, such as
$Cu(C_{6}D_{5}COO)_{2}3D_{2}O$\cite{Dender1,Dender2},
$Yb_{4}As_{3}$\cite{Kohgi,Fulde,Oshikawa},
$BaCu_{2}Si_{2}O_{7}$\cite{Tsukada}, $\alpha-Fe_{2}O_{3}$,
$LaMnO_{3}$\cite{Grande} and $K_{2}V_{3}O_{8}$\cite{Greven},
exhibit unusual and interesting magnetic properties due to
quantum fluctuations and/or in the presence of an applied magnetic
field\cite{Grande,Yildirim,Katsumata}. $La_{2}CuO_{4}$ also
belongs to the class of DM antiferromagnets, which is a parent
compound of high-temperature superconductors\cite{Kastner}. This
has stimulated extensive investigations of the properties which are created
from DM interaction. This interaction is however, rather difficult
to handle analytically, which makes  the interpretation of
experimental data to be hard. In addition, more knowledge in this respect expand our understanding of many
interesting quantum phenomena of low-dimensional magnetic materials.

Recent discovery of an unusual strong coupling between the
ferroelectric (FE) and magnetic order parameters has also revived
the interest in the magnetoelectric effect\cite{Fiebig}. Due to
the possibility of easily controlling the electrical properties
using magnetic field, search of compounds, in which the magnetic
order is incommensurate with lattice period, is of particular
interest for future applications\cite{Kimura1,Hur}. Generally,
certain types of magnetic order can lower the symmetry of the
system to that of the polar groups, which allow for
ferroelectricity. According to the recent experimental results,
helical magnetic structure are the most likely candidates to host
ferroelectericity\cite{Kimura2,Lawes,Kenzelmann}. It has been
shown that the DM interaction induces a FE lattice displacement
and helps to stabilize helical magnetic structures at low
temperature\cite{Sergienko}.

In the present paper, we have studied a one dimensional AF Ising
model with DM interaction using the quantum renormalization group
(QRG) and numerical exact diagonalization methods.
In the next section the QRG approach will be
explained and the renormalization of coupling constants are
obtained. In section III, we will obtain the phase diagram, fixed
points, critical points and the staggered magnetization
as the order parameter of the model. We will also introduce the
chiral order as an ordering which is created by DM interaction.
The exponent which shows the divergence of correlation function
close to the critical point $(\nu)$, the dynamical exponent $(z)$
and the exponent which shows the vanishing of staggered
magnetization near the critical point $(\beta)$ will be also
calculated. We then present the numerical exact diagonalization results
on finite sizes of $N=12, 16, 20$ and $24$.
In section IV we will calculate the renormalization of
entanglement\cite{kargarian2,kargarian1} for this model and we
will show that it has a scaling behavior near the QCP which is
directly related to critical properties of the model.

%%%%%%%%%%%%%%%%%%%%%%%%%%%%%%%%%%%%%%%%%%%%%%%%%%%%%%%%%%%%%%%%%%%%%%%%%%%%%

\section{Quantum renormalization group \label{qrg}}

%%%%%%%%%%%%%%%%%%%%%  Fig.1   %%%%%%%%%%%%%%%%%%%%%%%
\begin{figure}
\begin{center}
\includegraphics[width=8cm]{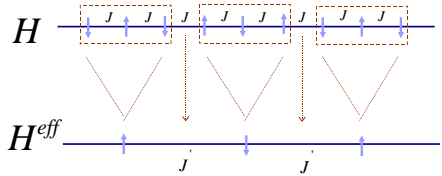}
\caption{(color online)The decomposition of chain into three site
blocks Hamiltonian ($H^{B}$) and inter-block Hamiltonian
($H^{BB}$).} \label{fig1}
\end{center}
\end{figure}
%%%%%%%%%%%%%%%%%%%%%%%%%%%%%%%%%%%%%%%%%%%%%%%%%%%%%%%

The main idea of the RG method is the mode elimination or the
thinning of the degrees of freedom followed by an iteration which
reduces the number of variables step by step until a more
managable situation is reached. We have implemented the
Kadanoff's block approach to do this purpose, because it is well
suited to perform analytical calculations in the lattice models
and they are conceptually easy to be extended to the higher
dimensions\cite{miguel1,miguel-book,Langari,Jafari}. In the
Kadanoff's method, the lattice is divided into blocks in which the
Hamiltonian is exactly diagonalized. By selecting a number of
low-lying eigenstates of the blocks the full Hamiltonian is
projected onto these eigenstates which gives the effective
(renormalized) Hamiltonian.

The Hamiltonian of Ising model with DM interaction in the $z$
direction on a periodic chain of $N$ sites is

\bea \label{eq1} H=\frac{J}{4}
\Big[\sum_{i=1}^{N}\sigma_{i}^{z}\sigma_{i+1}^{z}+D(\sigma_{i}^{x}\sigma_{i+1}^{y}-\sigma_{i}^{y}\sigma_{i+1}^{x})\Big]
\eea

The effective Hamiltonian (in the first order renormalization group prescription) is
\bea \no H^{eff}=H^{eff}_{0}+H^{eff}_{1},~~~~~~~~~\\
\no
H^{eff}_{0}=P_{0}H^{B}P_{0}~~~,~~~H^{eff}_{1}=P_{0}H^{BB}P_{0}.
\eea We have considered a three site block procedure defined in
Fig.(\ref{fig1}). The block Hamiltonian $(H_{B}=\sum h_{I}^{B})$
of the three sites and its eigenstates and eigenvalues are given
in Appendix A. The three site block Hamiltonian has four doubly
degenerate eigenvalues (see appendix A). $P_{0}$ is the projection
operator of the ground state subspace which is defined by
$\big(P_{0}=|\Uparrow\rangle\langle\psi_{0}|+|\Downarrow\rangle\langle\psi_{0}'|\big)$,
where $|\psi_{0}\rangle$ and $|\psi_{0}'\rangle$ are the doubly
degenerate ground states, $|\Uparrow\rangle$ and
$|\Downarrow\rangle$ are the renamed base kets in the effective
Hilbert space. We have kept two states ($|\psi_{0}\rangle$ and
$|\psi_{0}'\rangle$) for each block to define the effective (new)
sites. Thus, the effective site can be considered as a spin 1/2.
The effective Hamiltonian is not exactly similar to the initial
one, i.e, the sign of DM interaction is changed \bea \no
H^{eff}=\frac{J^{'}}{4}
\Big[\sum_{i=1}^{N}\sigma_{i}^{z}\sigma_{i+1}^{z}-D^{'}(\sigma_{i}^{x}\sigma_{i+1}^{y}-\sigma_{i}^{y}\sigma_{i+1}^{x})\Big],
\eea
where $J'$ and $D'$ are the renormalized coupling constants.
To have a self-similar Hamiltonian, we implement a $\pi$ rotation
around $x$ axis on all sites $(\sigma^{z}_{i}\rightarrow
-\sigma^{z}_{i}~,~\sigma^{y}_{i}\rightarrow -\sigma^{y}_{i})$. We
note to interpret our final results in terms of this
transformation. The renormalized coupling constants are functions
of the original ones which are given by the following equations.
\bea \no J'=J(\frac{1+q}{2q})^{2}~~,~~
D'=\frac{16D^{3}}{(1+q)^{2}}~~~,~~~q=\sqrt{1+8D^{2}}. \eea
We will
implement this approach in the next sections to obtain the phase
diagram and entanglement properties of the model.

%%%%%%%%%%%%%%%%%%%%%%%%%%%%%%%%%%%%%%%%%%%%%%%%%%%%%%%%%%%%%%%%%
\section{Phase Diagram}

The RG equations show the scaling of $J$ coupling to zero which
represents the renormalization of energy scale. At zero
temperature, a phase transition occurs upon variation of the
parameters in the Hamiltonian. In the absence of DM interaction
$(D=0)$ the ground state of the Ising model is the Neel ordered
state. However, for $D\neq0$ the DM interaction makes a tendency
for spins to be oriented in the $XY$ plane. A nonzero value of $D$
increases the fluctuations which destroys the AF ordering in $z$
direction at some finite value of $D=D_c$. Simultaneously, the
chiral order grows up and will saturate as $D\rightarrow\infty$.
The quantum phase transition can be interpreted as the
antiferromagnet to saturated chiral (SC) order transition at
$D=D_c$. The RG flow shows that in the AF phase $(D<D_c=1)$, the
DM coupling $(D)$ goes to zero and in the SC phase $D$ goes to
infinity (Fig.(\ref{fig2})). We have probed the AF-SC transition
by calculating the staggered magnetization $S_{M}$ (See appendix
B) in the $z$-direction as an order parameter (Fig.(\ref{fig3})),
\be \label{sm} 
S_M=\frac{1}{N}\sum_{i=1}^{N} \frac{(-1)^i}{2}
\langle \sigma_i^z  \rangle. \ee 
$S_M$ is zero in the SC phase and
has a nonzero value in the AF phase. Thus, the staggered
magnetization is the proper order parameter to represent the AF-SC
transition. We have plotted $S_{M}$ versus $D$ in
Fig.(\ref{fig3}). It has its maximum value at $D=0$ and
continuously decreases with increase of $D$ to zero at $D=1$.
Moreover, we have calculated the chiral order\cite{Kawamura}
($C_{h}$)  in the $z$ direction (See appendix B) which increases
with $D$ and saturates for $D\rightarrow\infty$ (Fig.\ref{fig3}),
\bea \no
C_{h}=\frac{1}{N}\sum_{i=1}^{N}\frac{1}{4}\langle(\sigma_{i}^{x}\sigma_{i+1}^{y}-\sigma_{i}^{y}\sigma_{i+1}^{x})\rangle.
\eea The chiral order has a nonzero value in both AF and SC phases
which can not be a proper order parameter to distinguish the
quantum phase transition. However, it shows that the onset of DM
interaction sets up the chiral order  immediately. A classical
picture of the chiral order in terms of the spin projection on the
$xy$-plane has been plotted in Fig.\ref{fig4}.

We have also calculated the critical exponents at the critical point
$(D=1)$. In this respect, we have obtained the dynamical exponent,
the exponent of order parameter and the diverging exponent of the
correlation length. This corresponds to reaching the critical point
from the AF phase by approaching $D\rightarrow 1$. The dynamical
exponent is given by $z=[ln(J/J')_{D=1}]/[ln(n_B)] \simeq 0.73$, where
$n_B=3$ is the number of sites in each block.
The staggered magnetization close to the quntum critical point 
goes to zero like $S_M \sim |D -1|^{\beta}$ where $\beta \simeq 1.15$ and is obtained 
by $\beta=[ln(S'_M/S_M)]/ln[\frac{d D'(D)}{d D}]|_{D=1}$ where prime denotes the
renormalized quantity. The correlation length diverges $\xi \sim |D -1|^{-\nu}$ with exponent
$\nu\simeq2.15$ which is expressed by 
$\nu=[ln(n_B)]/ln[\frac{d D'(D)}{d D}]|_{D=1}$.
The detail of this calculation is similar to what
has been presented in Ref.[\onlinecite{miguel1}].

%%%%%%%%%%%%%%%%%%%%%  Fig.2   %%%%%%%%%%%%%%%%%%%%%%%
\begin{figure}
\begin{center}
\includegraphics[width=8cm]{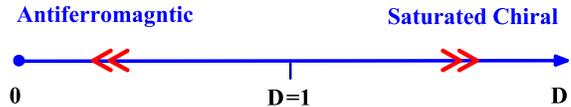}
\caption{(color online) Phase diagram of the Ising model with DM
interaction. Arrows show the running of coupling constant under
RG iteration.} \label{fig2}
\end{center}
\end{figure}
%%%%%%%%%%%%%%%%%%%%%%%%%%%%%%%%%%%%%%%%%%%%%%%%%%%%%%%

%%%%%%%%%%%%%%%%%%%%%  Fig.3   %%%%%%%%%%%%%%%%%%%%%%%
\begin{figure}
\begin{center}
\includegraphics[width=8cm]{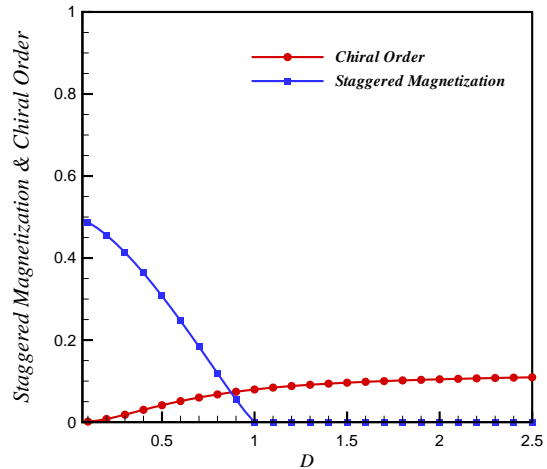}
\caption{(color online) Chiral order (filled circles) and Staggered Magnetization (filled squares)
versus D.} \label{fig3}
\end{center}
\end{figure}
%%%%%%%%%%%%%%%%%%%%%%%%%%%%%%%%%%%%%%%%%%%%%%%%%%%%%%%

%%%%%%%%%%%%%%%%%%%%%  Fig.4   %%%%%%%%%%%%%%%%%%%%%%%
\begin{figure}
\begin{center}
\includegraphics[width=8cm]{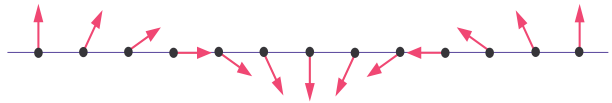}
\caption{(color online) A classical picture of spin orientation in the $xy$ plain where the
angle between neibouring spins depend on the D value (see appendix \ref{classical}).} \label{fig4}
\end{center}
\end{figure}
%%%%%%%%%%%%%%%%%%%%%%%%%%%%%%%%%%%%%%%%%%%%%%%%%%%%%%%

\section{Numerical results}\label{numerics}
\begin{figure}
\begin{center}
\includegraphics[width=8cm]{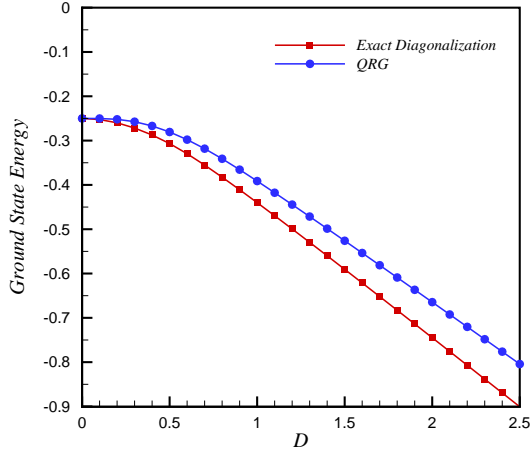}
\caption{(color online) The ground state energy versus $D$. The
result of QRG (filled circle) is compared with the extrapolated
values of exact diagonalization (filled squares).} \label{fig5}
\end{center}
\end{figure}

We have implemented the exact diagonalization method based on
Lanczos algorithm to get the ground state properties of the
Hamiltonian defined in Eq.(\ref{eq1}). The Hamiltonian does not
commute with $S^z=(1/2)\sum_i \sigma^z_i$, which imposes to
consider the full Hilbert space for computations. We have
considered a periodic chain of length $N=8, 12, 16, 20$ and $24$
in our calculations. We have first calculated the ground state
energy for different sizes. We have observed the size dependence
for ground state energy is weak which make us to extrapolate our
results to get the ground state energy per site ($E_0/N$) as its
thermodynamic limit ($N\rightarrow \infty$). We have plotted
$E_0/N$ versus $D$ in Fig.(\ref{fig5}). Our results show that the
QRG result for $E_0$ is close to the exact diagonalization one
which justifies the correct trend versus $D$ although the values
have around 10 percent error.  It is a good evidence that the QRG
result is reliable at least to get the qualitative picture of the
model.

%%%%%%%%%%%%%%%%%%%%%  Fig.5   %%%%%%%%%%%%%%%%%%%%%%%

%%%%%%%%%%%%%%%%%%%%%%%%%%%%%%%%%%%%%%%%%%%%%%%%%%%%%%%

%%%%%%%%%%%%%%%%%%%%%  Fig.6   %%%%%%%%%%%%%%%%%%%%%%%
\begin{figure}[<t>]
\begin{center}
\includegraphics[width=8cm]{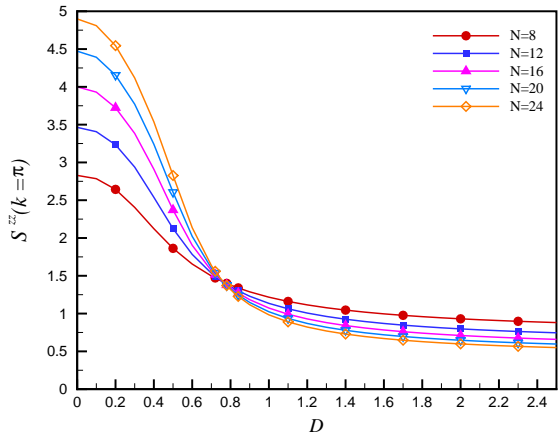}
\caption{(color online) The z-component static structure factor at antiferromagnetic ordering vector $k=\pi$ versus D.
A more clear picture about the crossing point is plotted in Fig.(\ref{fig8}).} \label{fig6}
\end{center}
\end{figure}

%%%%%%%%%%%%%%%%%%%%%%%%%%%%%%%%%%%%%%%%%%%%%%%%%%%%%%%
%%%%%%%%%%%%%%%%%%%%%  Fig.8   %%%%%%%%%%%%%%%%%%%%%%%
\begin{figure}[<t>]
\begin{center}
\includegraphics[width=8cm]{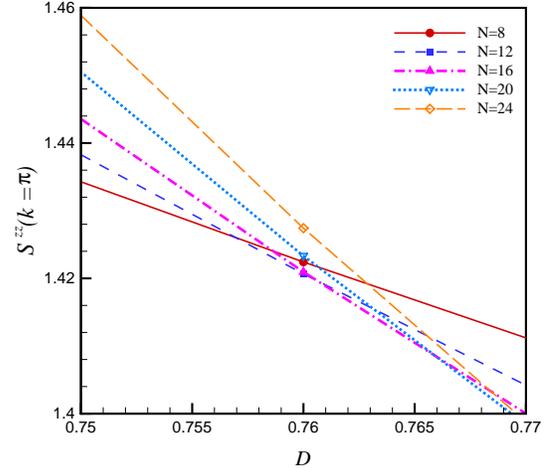}
\caption{(color online) A closer look at the crossing point of Fig.(\ref{fig6}) which shows that different plots
do not cross each other at a single point. The crossing point of two successive size occurs at larger value
of D upon increasing size.} \label{fig8}
\end{center}
\end{figure}

%%%%%%%%%%%%%%%%%%%%%%%%%%%%%%%%%%%%%%%%%%%%%%%%%%%%%%%

%%%%%%%%%%%%%%%%%%%%%  Fig.7   %%%%%%%%%%%%%%%%%%%%%%%
\begin{figure}[<t>]
\begin{center}
\includegraphics[width=8cm]{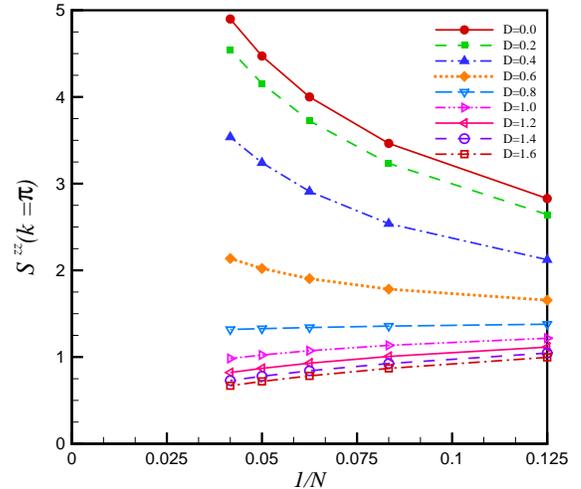}
\caption{(color online) The z-component structure factor for different $D$ versus $1/N$. Plots for $D\lesssim0.8$
show diverging behaviour as $N\rightarrow \infty$ while those for $D>0.8$ becomes finite in the thermodynamic
limit.} \label{fig7}
\end{center}
\end{figure}

%%%%%%%%%%%%%%%%%%%%%%%%%%%%%%%%%%%%%%%%%%%%%%%%%%%%%%%

Since the numerics is done on finite size systems the symmetry
breaking can not occur in our calculation to show the nonzero
value of the order parameter. Instead, the structure factor shows
a divergent bahaviour at ordering momentum by increasing the size
of system. The structure factor at momentum $k$ is defined by the
following relation \be \label{sk} S^{\alpha
\alpha}(k)=\frac{1}{\sqrt{N}}\sum_{r=0}^{N-1} \langle
\sigma_i^{\alpha} \sigma_{i+r}^{\alpha} \rangle e^{i k r}. \ee The
$z$-component of structure factor ($S^{zz}(k)$) versus $k$ has a
sharp peak at $k=\pi$  for $D\lesssim0.8$ representing the
antiferromagnetic order. To justify if the peak corresponds to the
true long range order (LRO) or it is just a local order we have
computed the structure factor for different size of chains ($N$).
We have plotted in Fig.(\ref{fig6}) $S^{zz}(k=\pi)$ versus $D$ for
different chain lengths ($N$). The peak height increases for
$D\lesssim0.8$ and decreases for $D>0.8$. To have a clear picture
of these data, we have plotted in Fig.(\ref{fig7}) the same data
for fixed $D$ value versus $1/N$. We observe a divergent behaviour
for the $z$-component structure factor for $D\lesssim0.8$ and
diminishing for $D>0.8$. This justifies antiferromagnetic LRO for
$D\lesssim0.8$. Although the plots for different $N$ in
Fig.(\ref{fig6}) show to cross each other at a single point
$D^{*}=0.76$ the fine tunning data close to this point
(Fig.(\ref{fig8})) represent different crossings for two
successive $N$ values. It is the manifestation of finite size
scaling which exist in our numerics. Therefore, the true critical
point ($D_c$) which should be the case for $N\rightarrow \infty$
is greater than $D^{*}=0.76$. The investigation which shows the
relation between the Ising model and DM interaction with the
anisotropic Heisenberg model (XXZ) verifies that the critical
point should be at $D_c=1$. It will be discussed in
Sec.\ref{conclusion}.

\begin{figure}[<t>]
\begin{center}
\includegraphics[width=8cm]{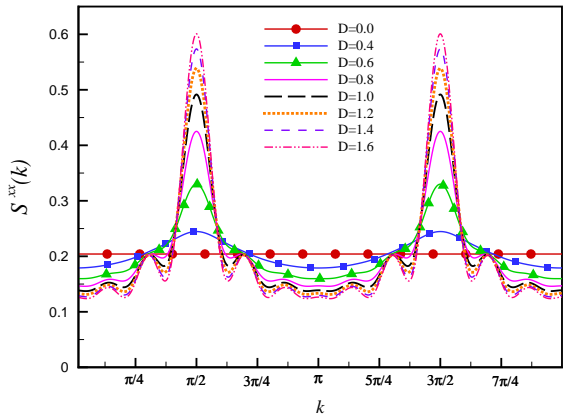}
\caption{(color online) The x-component structure factor versus momentum ($k$)  for different $D$.} \label{fig9}
\end{center}
\end{figure}

\begin{figure}[<t>]
\begin{center}
\includegraphics[width=8cm]{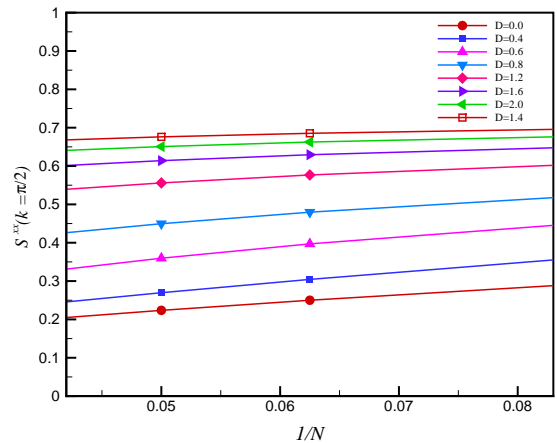}
\caption{(color online) The finite size scaling of $S^{xx}(k=\pi/2)$. All plot show non-divergent behaviour
as $N\rightarrow \infty$ representing no long range order but a local order.} \label{fig10}
\end{center}
\end{figure}

To get a picture on the type of ordering in the $xy$ plane we have
plotted the $x$ structure factor of $N=24$ for different $D$
values versus $k$ in Fig.(\ref{fig9}). Due to symmetry we have
$S^{xx}(k)=S^{yy}(k)$, thus, we only present data for $S^{xx}(k)$.
The $x$-component structure factor show two strong peaks at
$k=\pi/2, 3\pi/2$. It is a justification of spiral order in the
$xy$ plane. However, this is a local ordering and is not a true
LRO. We have also plotted in Fig.(\ref{fig10}) the value of
$S^{xx}(k=\pi/2)$ versus $1/N$ for different $D$ values. All data
in Fig.(\ref{fig10}) show diminishing behaviour as $N\rightarrow \infty$.
This justifies that the spiral (chiral) order which exist in the
$xy$ plane for $D\neq 0$ is not a true LRO.

%%%%%%%%%%%%%%%%%%%%%  Fig.9   %%%%%%%%%%%%%%%%%%%%%%%

%%%%%%%%%%%%%%%%%%%%%%%%%%%%%%%%%%%%%%%%%%%%%%%%%%%%%%%

%%%%%%%%%%%%%%%%%%%%%  Fig.10   %%%%%%%%%%%%%%%%%%%%%%%

%%%%%%%%%%%%%%%%%%%%%%%%%%%%%%%%%%%%%%%%%%%%%%%%%%%%%%%

\section{Entanglement and its scaling property \label{ent-rg}}
%\section{mehdi-scaling in the Ising+DM model}

In this section we calculate the entanglement of the model using
the idea of renormalization group \cite{kargarian2}. As we have
mentioned previously, a finite size block is treated exactly to
calculate the physical quantities. The coupling constants of a
finite size blocks are renormalized via the QRG prescription to
give the large size behavior. Bipartite entanglement, i.e the
entanglement between some degrees of freedom and the rest of
system, is quantified by von-Neumann entropy of eigenvalues of the
reduced density matrix. In our case, we first calculate the
entropy of the middle site and the remaining sites  of a single
block (see Fig.1). The entanglement is easily calculated, since
the density matrix is defined by

\be \label{eq35} \varrho=|\psi_{0}\rangle\langle\psi_{0}|, \ee

where $|\psi_{0}\rangle$ has been introduced in Appendix (Eq.(\ref{eqA1})).
The results will be the same if we consider $|\psi'_{0}\rangle$ to
construct the density matrix.

%%%%%%%%%%%%%%%%%%%%%  Fig.11   %%%%%%%%%%%%%%%%%%%%%%%
\begin{figure}
\begin{center}
\includegraphics[width=8cm]{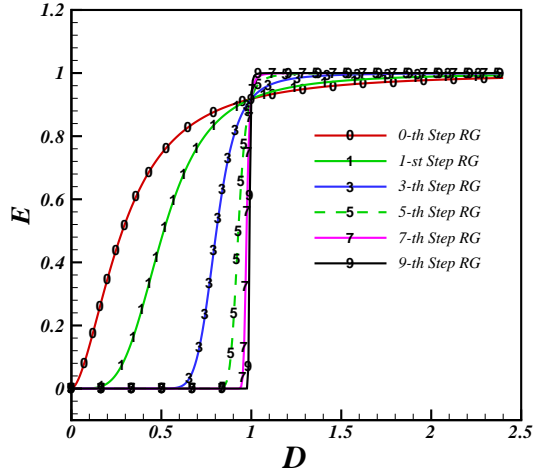}
\caption{(color online) Representation of the evolution of
entanglement entropy in terms of RG iterations.} \label{fig11}
\end{center}
\end{figure}
%%%%%%%%%%%%%%%%%%%%%%%%%%%%%%%%%%%%%%%%%%%%%%%%%%%%%%%

The density matrix defined in Eq.(\ref{eq35}) is traced over sites 1
and 3 to get the reduced density matrix for site 2 ($\varrho_{2}$)
which gives
\bea \label{eq36} \varrho_{2}=\frac{1}{2q(1+q)} \left(
\begin{array}{cc}
8D^{2} & 0 \\
0 & (1+q)^{2} \\
\end{array}
\right). \eea
The von-Neumann entropy is then
\bea \label{eq44}
E=-\frac{8D^{2}}{2q(1+q)}\log_{2}\frac{8D^{2}}{2q(1+q)} \nonumber \\
-\frac{(1+q)^{2}}{2q(1+q)}\log_{2}\frac{(1+q)^{2}}{2q(1+q)}. \eea

In the spirit of RG, the first iteration of RG represents a chain
of $3^{2}$ sites  which is described effectively by  three
effective sites interacting via the renormalized coupling
constants. Having this in mind, we understand that in the first RG
iteration the von-Neumann entropy with renormalized coupling
constant yields the entanglement between effective degrees of
freedom. The variation of entanglement ($E$) versus $D$ is plotted
in Fig.\ref{fig11}. Different plots show the evolution of $E$
under QRG iterations. In other words, the different iteration of
QRG show how the entanglement evolves as the size of chain is
increased. Long wavelength behaviors are captured as the RG
iterations are increased. In Fig.\ref{fig11} we see that in the
gapped phase, i.e AF, and the long-wavelength limit the
entanglement is suppressed while in the SC phase the entanglement
gets maximum value due to the DM interaction in the $XY$ plane
that induces a state with strong quantum correlation. Such a
behavior has also be seen in the $XXZ$ model\cite{kargarian1}.

%%%%%%%%%%%%%%%%%%%%%  Fig. 12  %%%%%%%%%%%%%%%%%%%%%%%
\begin{figure}
\begin{center}
\includegraphics[width=8cm]{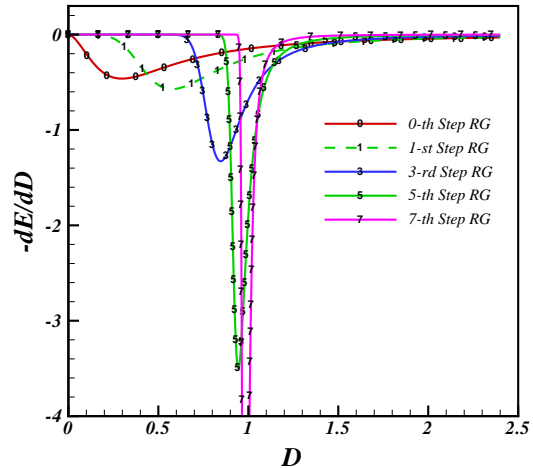}
\caption{(color online) First derivative of entanglement entropy
and its manifestation towards divergence as the number of RG
iterations increases (Fig.\ref{fig11}).} \label{fig12}
\end{center}
\end{figure}
%%%%%%%%%%%%%%%%%%%%%%%%%%%%%%%%%%%%%%%%%%%%%%%%%%%%%%%%%%%%
%%%%%%%%%%%%%%%%%%%%%  Fig. 13   %%%%%%%%%%%%%%%%%%%%%%%
\begin{figure}
\begin{center}
\includegraphics[width=8cm]{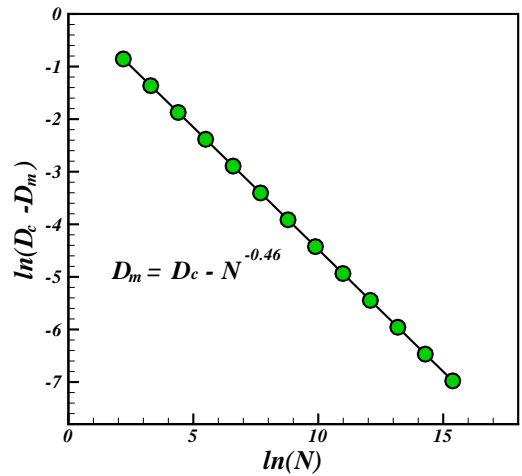}
\caption{(color online) The scaling behavior of $D_{m}$ in terms
of system size ($N$) where $D_{m}$ is the position of minimum in
Fig.\ref{fig12}.} \label{fig13}
\end{center}
\end{figure}
%%%%%%%%%%%%%%%%%%%%%%%%%%%%%%%%%%%%%%%%%%%%%%%%%%%%%%%%%%%%

%%%%%%%%%%%%%%%%  Fig.14  %%%%%%%%%%%%%%%%%%%%%%%
\begin{figure}
\begin{center}
\includegraphics[width=8cm]{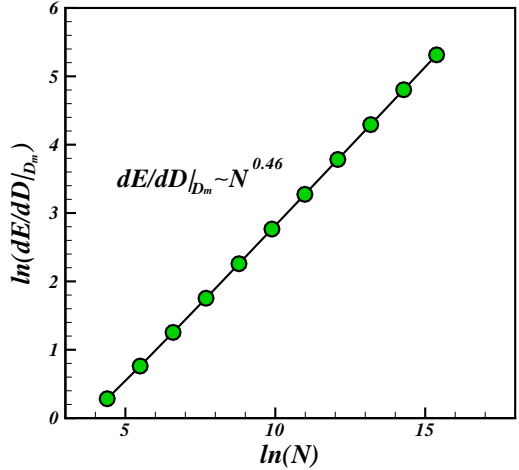}
\caption{(color online) The logarithm of the absolute value of
minimum, $\ln(dE/dD\mid_{D_{m}})$, versus the logarithm of chain
size, $\ln(N)$, which is linear and shows a scaling behavior. Each
point corresponds to the minimum value of a single plot of
Fig.\ref{fig12}.} \label{fig14}
\end{center}
\end{figure}

%%%%%%%%%%%%%%%%%%%%%%%%%%%%%%%%%%%%%%%%%%%%%%%%%%%%%%%

A common feature of the second order phase transitions is the
appearance of nonanalytic behavior in some physical quantities or
their derivatives as the critical point is crossed\cite{goldenfeld}.
It is also accompanied by a scaling behavior since the correlation
length diverges and there is no characteristic length scale in the
system at the critical point. Entanglement as a direct measure of
quantum correlations indicates the critical behavior such as
diverging of its derivative  as the phase transition is crossed
\cite{Osterloh}. It  has been verified that the entanglement in the
vicinity of critical point of Ising model in transverse field (ITF)
and $XX$ model in transverse field shows a scaling behavior.
Investigating the nonanaliticity, e.g a divergence, and finite size
scaling provides excellent estimates for the quantum critical point.
A precise connection between the entanglement in quantum information
theory and the critical phenomena in condensed matter physics has
been established\cite{vidal}, where the scaling properties of the
entanglement in spin chain systems, both near and at a quantum
critical point have been investigated. The first derivative of
entanglement let us to get more insight on the qualitative variation
of the ground state as the critical point is touched. To this end we
have calculated the first derivative of entanglement which has been
depicted in Fig.\ref{fig12}. Such a computation determines the
scaling law of entanglement in one-dimensional spin systems, while
explicitly uncovering an accurate correspondence with the critical
properties of the model. As the size of system becomes large through
RG iterations, the derivative of entanglement tends to diverge close
to the critical point. All plots in Fig.\ref{fig12} with respect to
the critical point have an asymmetrical shape. Each plot  reveals a
minimum in the gapped phase, i.e AF for $0\leq D <1$, the minimum
becomes more pronounced close to the critical point, $D=1$. It
manifests that the ground state of the gapped phase of the model
undergoes a strong qualitative change when approaching the quantum
critical point while the corresponding change in the SC phase is
rather small. A similar situation has also been observed in the
$XXZ$ model \cite{kargarian1}. This behavior is comparable with
results on ITF model where the system in both sides of the critical
point is gapfull, so the derivative of the entanglement tends to
diverge symmetrically \cite{kargarian2}.

More information can be obtained when the minimum values  of each
plot and their positions are analyzed. The position of the minimum
($D_m$) of $\frac{dE}{dD}$ tends towards the critical point like
$D_{m}=D_{c}-N^{-0.46}$ which has been plotted in Fig.\ref{fig13}.
Moreover, we have derived the scaling behavior of
$y\equiv|\frac{dE}{dD}|_{D_m}$ versus $N$. This has been plotted
in Fig.\ref{fig14} which shows a linear behavior of $ln(y)$ versus
$ln(N)$. The exponent for this behavior is
$\mid\frac{dE}{dD}|_{D_m} \sim N^{0.46}$. This results justify
that the RG implementation of entanglement  truly capture the
critical behavior of the model at $D=1$. It should be emphasized
this exponent is directly related to the correlation length
exponent, $\nu$, close to the critical point. It has been shown in
Ref.[\onlinecite{kargarian1}] that
$\mid\frac{dE}{dD}\mid_{D_{c}}\sim N^{1/\nu}$  and
$D_{m}=D_{c}+N^{-1/\nu}$.

\begin{figure}
\begin{center}
\includegraphics[width=8cm]{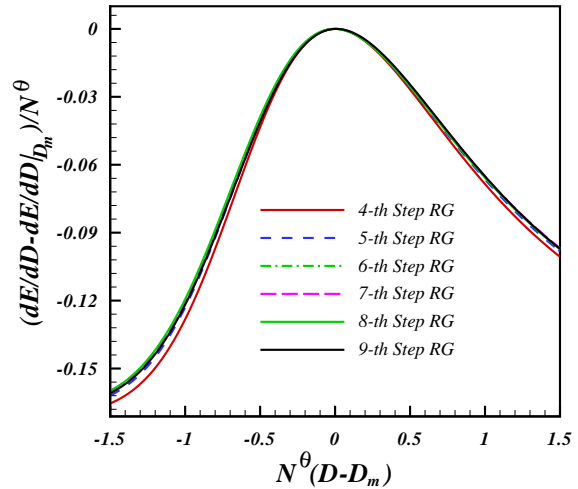}
\caption{(color online) The finite size scaling is performed via
the RG treatment for the power-law scaling. Each curve corresponds
to a definite size of the system, i.e $N=3^{n+1}$. The exponent
$\theta$ is ascribed to the correlation length critical exponent
$\nu$ via $\theta=1/\nu$ } \label{fig15}
\end{center}
\end{figure}

To study the scaling behavior of the entanglement entropy around
the critical point, we perform finite scaling analysis. Since the
minimum value of derivative of entanglement entropy scales
power-law. According to the scaling ansatz, the rescaled
derivative of entanglement entropy around its minimum value,
$D_{m}$, is just a function of rescaled driving parameter like:
\bea \no
\frac{\frac{dE}{dD}-\frac{dE}{dD}|_{D_{m}}}{N^{\theta}}=F[N^{\theta}(D-D_{m})]
\eea where, $F(x)$ is a universal function that does not deponed
on the system size, and the exponent $\theta$ is just the inverse
of the critical exponent $\nu$, i.e $\theta=1/\nu$. The
manifestation of the finite size scaling is shown in
Fig.\ref{fig15}. It is clear that the different curves which are
resemblance of various system sizes collapse to a single universal
curve. It must be noticed that the $n-th$ RG iteration describes a
system with $3^{n+1}$ sites which is effectively represented by a
three site model through the RG treatment.

%%%%%%%%%%%%%%%%  Fig.15  %%%%%%%%%%%%%%%%%%%%%%%
%%%%%%%%%%%%%%%%%%%%%%%%%%%%%%%%%%%%%%%%%%%%%%%%%%%%%%%
%%%%%%%%%%%%%%%%%%%%%%%%%%%%%%%%%%%%%%%%%%%%%%%%%%%%%%%%%%%%%%%%%%%%%%%%%%%%%%%%%%%%%%%%%%%%%%%%%%%%%%%%%%%%%%

\section{Summary and conclusions \label{conclusion}}

We have applied the quantum RG approximation to obtain the phase diagram,
staggered magnetization, chiral order and the entanglement
properties of Ising model with DM interaction. Tunning the DM
interaction dictated the system to fall into different phases, i.e
antiferromagnetic phase with nonzero staggered magnetization (as order parameter) and chiral
one with vanishing order parameter. The critical point of the phase transition is at
$D_c=1$ where the quantum fluctuations have dominant effect which
arises from the DM interaction and eventually destroy the order in
the AF phase. Although the DM interaction drives the spins to
leave their ordering in the z-direction, i.e staggered
magnetization, a chiral saturated phase has been arisen,
Fig.\ref{fig2}.

The numerical exact diagonalization which has been done on finite
sizes ($N=12, 16, 18, 20$) justifies the results of QRG. We have
obtained a fairly well agreement in the ground state energy of QRG
approach with exact diagonalization method. The divergence of
$z$-component structure factor at momentum $k=\pi$ when
$N\rightarrow \infty$ is the signature of antiferromagnetic long
range order for $D<D_c$. The value of critical point which is read
from exact diagonalization is $D_c^{numeric}\simeq =0.8$ which is
$20$ percent different from what QRG gives,$D_c^{QRG}=1$. However,
we claim that the QRG result for $D_c$ should be more reliable as
will be discussed below in connection with anisotropic Heisenberg
(XXZ) model. Moreover, the size dependence of the critical point
($D_c$) is strong which demands the numerics on larger sizes and
also a scaling analysis.

Besides, the entanglement entropy of the model at different RG
iterations was analyzed. As the long wavelength behavior of the
model is reached via the increasing of the RG iterations, the
entanglement entropy develops two distinct behavior proportional to
two different existing phases of the model. However nonanalytic
behavior close to the critical point of the model manifests itself
via the analysis of the first derivative of the entanglement
entropy. The divergence of the first derivative of the entanglement
entropy becomes more pronounced as long as the size of the system
becomes large in RG treatment. Critical point is touched by an
exponent which appears as the inverse of the critical exponent which
shows the divergence of the correlation length. Moreover, it is
found that as the critical point is touched from the gapped phase a
drastic change in the ground state occurs which manifests itself in
the evolution of the derivative of entanglement (see
Fig.\ref{fig12}). Such variation in the ground state structure also
appears in the $XXZ$ model. Finally finite size scaling reveals the
critical properties of the model is mirrored via the nonanalytic
behavior of the entanglement.

The one dimensional Ising model with DM interaction (Eq.(\ref{eq1})) is mapped
to the $XXZ$ chain via a nonlocal canonical transformation \cite{Aristov,Alcaraz},
\bea \label{ut}\no
U=\sum_{j=1}^{N}\alpha_{j}\sigma_{j}^{z}~~~&,&~~~\alpha_{j}=\sum^{j-1}_{m=1}
m \tan^{-1}(D), \\
 \no \tilde{\sigma}^{\pm}_{j}=e^{-iU}\sigma^{\pm}_{j}e^{iU}&,&~~~~~
\tilde{\sigma}^{z}_{j}=\sigma^z,\\
 \tilde{H}&=&e^{-iU} H e^{iU},
\eea
which gives
\be
\label{ht}
 \tilde{H} \sim \sum_i \big(\tilde{\sigma}^{x}_{i} \tilde{\sigma}^{x}_{i+1}+
\tilde{\sigma}^{y}_{i} \tilde{\sigma}^{y}_{i+1}+
(\frac{1}{D})\tilde{\sigma}^{z}_{i} \tilde{\sigma}^{z}_{i+1}\big).
\ee This transformation tells that the system is in the spin-fluid
phase for $D>1$ in terms of transformed spins. While, the model
represents the antiferromagnetic ($N\acute{e}el$) phase for $D<1$.
The spin-fluid phase ($D>1$) of the transformed Hamiltonian ($
\tilde{H}$) can be characterized by a string order parameter
\cite{Hida}. At the same time the saturated chiral phase ($D>1$)
of the original Hamiltonian (Eq.(\ref{eq1})) is represented by the
chiral order. It can be concluded that the chiral order
 with nonlocal spins is similar to the string order parameter. Therefore, it is suggested that the $XXZ$ model has
chiral order which is constructed with nonlocal spins. It will be
instructive to calculate the chiral order with nonlocal spins for
$XXZ$ model as a hidden order of the spin-fluid phase.

%%%%%%%%%%%%%%%%%%%%%%%%%%%%%%%%%%%%%%%%%%%%%%%%%%%%%%%%%%%%%%%%%
\begin{acknowledgments}
The authors would like to thank Prof. M. R. H. Khajehpour for careful
reading of the manuscript and fruitful discussions. We would like also acknowledge
J. Abouie for useful discussions.
This work was supported in part by the Center of Excellence in
Complex Systems and Condensed Matter (www.cscm.ir).
\end{acknowledgments}
%%%%%%%%%%%%%%%%%%%%%%%%%%%%%%%%%%%%%%%%%%%%%%%%%%%%%%%%%%%%%%%%

\appendix 

\section{The block Hamiltonian of three sites, its eigenvectors and
eigenvalues}

We have considered the three-site block (Fig.(\ref{fig1})) with the
following Hamiltonian

\bea \no h_{I}^{B}=\frac{J}{4}\Big[
(\sigma_{1,I}^{z}\sigma_{2,I}^{z}+\sigma_{2,I}^{z}\sigma_{3,I}^{z})+
D(\sigma_{1,I}^{x}\sigma_{2,I}^{y}\\
-\sigma_{1,I}^{y}\sigma_{2,I}^{x}+
\sigma_{2,I}^{x}\sigma_{3,I}^{y}-\sigma_{2,I}^{y}\sigma_{3,I}^{x})\Big].
\eea

The inter-block ($H^{BB})$ and intra-block ($H^{B}$) Hamiltonian for
the three sites decomposition are
%\begin{widetext}
\begin{eqnarray}
\no H^{B}&=&\frac{J}{4}\sum_{i=1}^{N/3}
\Big[\sigma_{1,I}^{z}\sigma_{2,I}^{z}+\sigma_{2,I}^{z}\sigma_{3,I}^{z}\\
\no &+&D(\sigma_{1,I}^{x}\sigma_{2,I}^{y}-
\sigma_{1,I}^{y}\sigma_{2,I}^{x}+
\sigma_{2,I}^{x}\sigma_{3,I}^{y}-\sigma_{2,I}^{y}\sigma_{3,I}^{x})\Big],\\
\no H^{BB}&=&\frac{J}{4}
\sum_{I=1}^{N/3}\Big[\sigma_{3,I}^{z}\sigma_{1,I+1}^{z}+
D(\sigma_{3,I}^{x}\sigma_{1,I+1}^{y}-\sigma_{3,I}^{y}\sigma_{1,I+1}^{x})\Big],
\end{eqnarray}
%\end{widetext}
where $\sigma_{j,I}^{\alpha}$ refers to the $\alpha$-component of
the Pauli matrix at site $j$ of the block labeled by $I$. The
exact treatment of this Hamiltonian leads to four distinct
eigenvalues which are doubly degenerate. The ground, first, second
and third excited state energies have the following expressions in
terms of the coupling constants.

\begin{eqnarray}
\no
|\psi_{0}\rangle&=&\frac{1}{\sqrt{2q(1+q)}}[2D|\downarrow\uparrow\uparrow\rangle+
i(1+q)|\uparrow\downarrow\uparrow\rangle-2D|\uparrow\uparrow\downarrow\rangle],\\
\no
|\psi_{0}'\rangle&=&\frac{1}{\sqrt{2q(1+q)}}[2D|\downarrow\downarrow\uparrow\rangle+
i(1+q)|\downarrow\uparrow\downarrow\rangle-2D|\uparrow\downarrow\downarrow\rangle],\\
\no
e_{0}&=&-\frac{J}{4}(1+q),\\
\no
|\psi_{1}\rangle&=&\frac{1}{\sqrt{2q(q-1)}}[2D|\downarrow\uparrow\uparrow\rangle-
i(q-1)|\uparrow\downarrow\uparrow\rangle-2D|\uparrow\uparrow\downarrow\rangle],\\
\no
|\psi_{1}'\rangle&=&\frac{1}{\sqrt{2q(q-1)}}[2D|\downarrow\downarrow\uparrow\rangle-
i(q-1)|\downarrow\uparrow\downarrow\rangle-2D|\uparrow\downarrow\downarrow\rangle],\\
\no
e_{1}&=&-\frac{J}{4}(1-q),\\
\no
|\psi_{2}\rangle&=&\frac{1}{\sqrt{2}}(|\uparrow\uparrow\downarrow\rangle+
|\downarrow\uparrow\uparrow\rangle)~~~,~~~
|\psi_{2}'\rangle=\frac{1}{\sqrt{2}}(|\downarrow\downarrow\uparrow\rangle+
|\uparrow\downarrow\downarrow\rangle),\\
\no
e_{2}&=&0,\\
\no |\psi_{3}\rangle&=&|\uparrow\uparrow\uparrow\rangle~~,~~
|\psi_{3}'\rangle=|\downarrow\downarrow\downarrow\rangle,\\
\label{eqA1} e_{3}&=&\frac{J}{2},
\end{eqnarray}

where $q$ is $q=\sqrt{1+8D^{2}}.$

$|\uparrow\rangle$ and $|\downarrow\rangle$ are the eigenstates of
$\sigma^{z}$. The projection operator is
\bea \no
P_{0}=|\Uparrow\rangle\langle\psi_{0}|+|\Downarrow\rangle\langle\psi_{0}'|.
\eea
The Pauli matrices in the effective Hilbert space have the following
transformations
%\begin{widetext}
\begin{eqnarray}
\no
P_{0}^{I}\sigma_{1,I}^{x}P_{0}^{I}&=&\frac{2D}{q}{\sigma'}_{I}^{y}~~,
~~P_{0}^{I}\sigma_{2,I}^{x}P_{0}^{I}=\frac{4D^2}{q(q+1)}{\sigma'}_{I}^{x}~~,\\
\no
P_{0}^{I}\sigma_{3,I}^{x}P_{0}^{I}&=&-\frac{2D}{q}{\sigma'}_{I}^{y}~~~~~,
~P_{0}^{I}\sigma_{1,I}^{y}P_{0}^{I}=-\frac{2D}{q}{\sigma'}_{I}^{x}~~,\\
\no
P_{0}^{I}\sigma_{2,I}^{y}P_{0}^{I}&=&\frac{4D^2}{q(q+1)}{\sigma'}_{I}^{y}~~,
~~P_{0}^{I}\sigma_{3,I}^{y}P_{0}^{I}=\frac{2D}{q}{\sigma'}_{I}^{x}~~,\\
\no
P_{0}^{I}\sigma_{1,I}^{z}P_{0}^{I}&=&\frac{1+q}{2q}{\sigma'}_{I}^{z}~~~~~~,
~P_{0}^{I}\sigma_{2,I}^{z}P_{0}^{I}=-\frac{1}{q}{\sigma'}_{I}^{z}~~,\\
\no
P_{0}^{I}\sigma_{3,I}^{z}P_{0}^{I}&=&\frac{1+q}{2q}{\sigma'}_{I}^{z}.
\end{eqnarray}
%\end{widetext}

\section{Order Parameter and Chiral Order}

\subsection{Staggered magnetization}

Generally, any correlation function can be calculated in the QRG
scheme. In this approach, the correlation function at each
iteration of RG is connected to its value after an RG iteration.
This will be continued to reach a controllable fixed point where
we can obtain the value of the correlation function. The staggered
magnetization in $\alpha$ direction can be written \bea
\label{eqB1} S_{M}=\frac{1}{N}\sum_{i}^{N}\langle
O|\frac{(-1)^i}{2}\sigma_{i}^{\alpha}|O\rangle, \eea where
$\sigma_{i}^{\alpha}$ is the Pauli matrix in the $i$th site and
$|O\rangle$ is the ground state of chain. The ground state of the
renormalized chain is related to the ground state of the original
one by the transformation, $P_{0}|O'\rangle=|O\rangle$. \bea \no
S_{M}=\frac{1}{N}\sum_{i}^{N}\langle
O'|P_{0}(\frac{(-1)^i}{2} \sigma_{i}^{\alpha})P_{0}|O'\rangle.
\eea This leads to the staggered configuration in the renormalized
chain. The staggered magnetization in $z$ direction is obtained

%\begin{widetext}
\begin{eqnarray}
\label{eqB2}
\no S_{M}^{0}&=&\frac{1}{N}\sum_{i=1}^{N}\langle
0| \frac{(-1)^i}{2} \sigma_{i}^{z}|0\rangle\\
\no
&=&\frac{1}{6}\frac{1}{\frac{N}{3}}\sum_{I=1}^{N/3}\Big[\langle
0'|P_{0}^{I}(-\sigma_{1,I}^{z}+
\sigma_{2,I}^{z}-\sigma_{3,I}^{z})P_{0}^{I}|0'\rangle\\
\no
&-&\langle0'|P_{0}^{I+1}(-\sigma_{1,I+1}^{z}+\sigma_{2,I+1}^{z}-\sigma_{3,I+1}^{z})P_{0}^{I+1}|0'\rangle\Big]\\
&=&-(\frac{2+q}{3q})\frac{1}{\frac{N}{3}}\sum_{I=1}^{N/3}\langle
0'| \frac{(-1)^I}{2} \sigma_{I}^{z}|0'\rangle=-\frac{\gamma^{0}}{3}S_{M}^{1},
\end{eqnarray}
%\end{widetext}
\\
where $S_{M}^{(n)}$ is the staggered magnetization at the $n$th step
of QRG and $\gamma^{(0)}$ is defined by $\gamma^{0}=(2+q)/q$.

%\be \label{eqB2} \gamma^{0}=\frac{2+q}{q} \ee

This process will be iterated many times by replacing $\gamma^{(0)}$
with $\gamma^{(n)}$. The expression for $\gamma^{(n)}$ is similar to
 $\gamma^{(0)}$ where the coupling constants should be replaced
by the renormalized ones at the corresponding RG iteration ($n$). The
result of this calculation has been presented in Fig.(\ref{fig3}).

\subsection{Chiral Order}
\begin{widetext}
\bea \no C_{h}&=&\frac{1}{N}\sum_{i=1}^{N}\frac{1}{4}\langle
0|(\sigma_{i}^{x}\sigma_{i+1}^{y}-\sigma_{i}^{y}\sigma_{i+1}^{x})|0\rangle\\
\no &=&\frac{1}{12}\frac{1}{\frac{N}{3}}\sum_{I=1}^{N/3}
\Big[\langle
0'|P_{0}(\sigma_{3,I}^{x}\sigma_{1,I+1}^{y}-\sigma_{3,I}^{y}\sigma_{1,I+1}^{x})P_{0}|0'\rangle
+\langle0'|P_{0}^{I}((\sigma_{1,I}^{x}\sigma_{2,I}^{y}-\sigma_{1,I}^{y}\sigma_{2,I}^{x})+
(\sigma_{2,I}^{x}\sigma_{3,I}^{y}-\sigma_{2,I}^{y}\sigma_{3,I}^{x}))P_{0}^{I}|0'\rangle\Big]\\
&=&\frac{1}{12}\frac{32D^{3}}{q^{2}(1+q)}+\frac{1}{3}\frac{1}{\frac{N}{3}}(\frac{2D}{q})^{2}
\sum_{I=1}^{N/3}\frac{1}{4}\langle
0'|(\sigma_{I}^{x}\sigma_{I+1}^{y}-\sigma_{I}^{y}\sigma_{I+1}^{x})|0'\rangle
\no =C^{0}+\frac{\Upsilon^{0}}{3}C^{1}_{h}~~,~~
C^{0}=\frac{1}{3}\frac{32D^{3}}{q^{2}(1+q)}~~,~~\Upsilon^{0}=(\frac{2D}{q})^{2}.
\eea
\end{widetext}

At the last step we use the following transformation,
$\sigma^{z}_{i}\rightarrow
-\sigma^{z}_{i}~,~\sigma^{y}_{i}\rightarrow -\sigma^{y}_{i}$.

%\bea \no
%C^{0}=\frac{1}{3}\frac{32D^{3}}{q^{2}(1+q)}~~~,~~~\Upsilon^{0}=(\frac{2D}{q})^{2}.
%\eea

\section{Classical Approximation}\label{classical}
In the classical approximation the spins are considered as classical vectors

%which form the spiral structure with a pitch angle $\varphi$
%between neighboring spins and canted angle $\theta$

\bea \no \sigma^{x}_{i}=\cos(i\varphi)\sin\theta~~,
\sigma^{y}_{i}=\sin(i\varphi)\sin\theta~~,
\sigma^{z}_{i}=\cos\theta,\eea

where $\varphi$ is the azimuthal angle measured from the $x$-axis
in the $xy$-plane and
$\theta$ is the polar angle measured down from the $z$-axis. The
classical energy per site for IDM Hamiltonian (Eq.(\ref{eq1})) is:
\bea \no
\frac{E_{cl}}{N}=\frac{J}{4}(\cos^{2}\theta+D\sin^{2}\theta\sin\varphi).
\eea
The minimization of classical energy with respect to the angles $\varphi$ and
$\theta$ shows that there are two different regions. (I) $D>1$, the minimum
of energy is obtained by $\theta=\frac{\pi}{2}$
and $\varphi=\arcsin(\frac{1}{D})$ which show the spins do not have a
projection on the $z$-axis and have the helical structure (see Fig.\ref{fig4})
in the $xy$ plain. In this region the minimum classical energy is
\bea \label{eqC1} \frac{E_{cl}^{I}}{N}=\frac{J}{4}. \eea

(II) $D<1$, the energy is minimized by
$\phi=\frac{\pi}{2}$ and arbitrary $\theta$ which corresponds to the
configuration with nonzero value of spins projection on $z$-axis
and helical structure of spins projection in the $xy$-plain where the
angle between spins are $\frac{\pi}{2}$. In this region the
minimum classical energy is
\bea \label{eqC2}
\frac{E_{cl}^{II}}{N}=\frac{J}{4}(\cos^{2}\theta+D\sin^{2}\theta).
\eea

One can see from Eq.(\ref{eqC1}) and Eq.(\ref{eqC2}) that the
transition between phase (I) and (II) takes place at $D=1$.


\section*{References}
\begin{thebibliography}{99}

\bibitem{Bardeen}
J. Bardeen, L. N. Cooper and J. R. Schrieffer, Phys. Rev.
\textbf{108}, 1175 (1957).

\bibitem{Laughlin}
R. B. Laughlin, Phys. Rev. Lett. \textbf{50}, 1395 (1983).

\bibitem{Sachdev}
S. Sachdev, Quantum phase transition, Cambridge Uni. Press (1999).

\bibitem{vojta}
M. Vojta, Rep. Prof. Phys. \textbf{66}, 2069 (2003) and references
therein.

\bibitem{Preskill}
J. Preskill, J. Mod. Opt \textbf{47}, 127 (2000).

\bibitem{Dzyaloshinskii}
I. Dzyaloshinskii, J. Phys. Chem. Solids \textbf{4}, 241 (1958).

\bibitem{Moriya}
T. Moriya, Phys. Rev \textbf{120}, 91 (1960).

\bibitem{Dender1}
D. C. Dender, D. Davidovic, D. H. Reich, C. Broholm, K. Lefmann,
and G. Aeppli, Phys. Rev. B \textbf{53}, 2583 (1996).

\bibitem{Dender2}
D. C. Dender, P. R. Hammar,D. H. Reich, C. Broholm , and G. Aeppli
Phys. Rev. Lett \textbf{79}, 1750 (1997).

\bibitem{Kohgi}
M. Kohgi, K. Iwasa, J. M. Mignot, B. Fak, P. Gegenwart, M. Lang,
A. Ochiai, H. Aoki, and T. Suzuki, Phys. Rev. Lett \textbf{86},
2439 (2001).

\bibitem{Fulde}
P. Fulde B. Schmidt, and P. Thalmeier, Europhys. Lett \textbf{31},
323 (1995).

\bibitem{Oshikawa}
M. Osikawa, K. Ueda, H. Aoki, A. Ochiai and M. Kohgi, J. Phys. Soc.
Jpn \textbf{68}, 3181 (1999); H. Shiba, K. Udea, and O. Sakai, J.
Phys. Soc. Jpn \textbf{69}, 1493 (2000)

\bibitem{Tsukada}
I. Tsukada, J. T. Takeya, T. Masuda and K. Uchinokura, Phys. Rev.
Lett \textbf{87}, 127203 (2001)

\bibitem{Grande}
b. Grande and Hk. M$\ddot{u}$ller-Buschbaum, Z. Anorg. Allg. Chem
\textbf{417}, 68 (1975).

\bibitem{Greven}
M. Greven, R. J. Birgeneau, Y. Endoh, M. A. Kastner, M. Matsuda, and
G. Shirane, Z. Phys. B \textbf{96}, 465 (1995).

\bibitem{Yildirim}
T. Yildirim, A. B. Harris, A. Aharony and O. Entin-Wohlman, Phys.
Rev. B. \textbf{52}, 10239 (1995).

\bibitem{Katsumata}
K. Katsumata, M. Hagiwara,Z. Honda, J. Satooka, A. Aharoy, R. J.
Birgeneau, F. C. Chou, O. E. Wohlman, A. B. Harris,M. A. Kastner, Y.
J. Kim, and Y. S. Lee, Europhys. Rev. Lett \textbf{54}, 508(2001).

\bibitem{Kastner}
M. A. Kastnetr, R. J. Birgeneau, G. Shirane and Y. Endoh, Rev. Mod.
Phys \textbf{70}, 897 (1998).

\bibitem{Fiebig}
M. Fiebig, J. Phys. D: Appl. Phys \textbf{38}, R123 (2005).


\bibitem{Kimura1}
T. Kimura and T. Goto, H. Shintani, K. Ishizaka, T. Arima, and Y.
Tokura Nature \textbf{426}, 55 (2003).

\bibitem{Hur}
N. Hur, S. Park, P. A. Sharma. J. S. Ahn, S. Guha, and S.-W. Cheong
Nature \textbf{429}, 392 (2004).

\bibitem{Kimura2}
T. Kimura, G. Lawes, and A. P. Ramirez, Phys. Rev. Lett \textbf{94},
137201 (2005).

\bibitem{Lawes}
G. Lawes, A. B. Harris, T. Kimura, N. Rogado, R. J. Cava, A.
Aharony, O. Entin-Wohlman, T. Yildrim, M. Kenzelmann, C. Broholm,
and A. P. Ramirez, Phys. Rev. Lett \textbf{95}, 087205 (2005).

\bibitem{Kenzelmann}
M. Kenzelmann, A. B. Harris, S. Jonas, C. Broholm, J. Schefer, S.
B. Kim, C. L. Zhang, S.-W. Cheong, O. P. Vajk, and J. W. Lynn,
Phys. Rev. Lett \textbf{95}, 087206 (2005).

\bibitem{Sergienko}
I. A. Sergienko and E. Dagotto, Phys. Rev. B \textbf{73}, 094434
(2006).

\bibitem{kargarian2}
M. Kargarian, R. Jafari, A. Langari, Phys. Rev. A \textbf{76},
060304(R) (2007).

\bibitem{kargarian1}
M. Kargarian, R. Jafari, A. Langari, Phys. Rev. A \textbf{77},
032346 (2008).


\bibitem{miguel1}
M. A. Martin-Delgado and G. Sierra, Int. J. Mod, Phys. A
\textbf{11}, 3145 (1996).


\bibitem{miguel-book}
G. Sierra and M. A. Martin Delgado, in \emph{Strongly Correlated
Magnetic and Superconducting Systems}, Lecture Notes in Physics Vo1.
478 (springer, Berlin, 1997).

\bibitem{Langari}
A. Langari, Phys. Rev. B \textbf{58}, 14467 (1998); \textbf{69},
100402(R) (2004).

\bibitem{Jafari}
R. Jafari, A. Langari, Phys. Rev. B \textbf{76}, 014412 (2007);
Physica A \textbf{364}, 213 (2006).


\bibitem{Kawamura}
H. Kawamura, Phys. Rev. B \textbf{38}, 4916 (1988).

\bibitem{goldenfeld}
N. Goldenfeld, \emph{Lectures on phase transition and the
Renormalization Group}, Addison-Wesley Publlishing group (1992).


\bibitem{Osterloh}
A. Osterloh, luigi Amico, G. Falci, and Rosario Fazio, Nature
\textbf{416}, 608 (2002).


\bibitem{vidal}
G. Vidal, J. I. Latorre, E. Rico, and A. Kitaev, Phys. Rev. Lett
\textbf{90}, 227902 (2003).


\bibitem{Aristov}
D. N. Aristov, S. V. Maleyev, Phys. Rev. B \textbf{62}, R751
(2000).

\bibitem{Alcaraz}
F. C. Alcaraz and W. F. Wreszinski, J. Stat. Phys. \textbf{58}, 45
(1990).

\bibitem{Hida}
K. Hida, Phys. Rev. B \textbf{45}, 2207, (1992).



\end{thebibliography}
\end{document}